\documentclass[11pt,onecolumn,english,preprintnumbers,amsmath,amssymb,floatfix,nofootinbib]{revtex4}
\usepackage{tikz,xcolor}
\usepackage[colorlinks = true,  
linkcolor = blue,
urlcolor  = blue,
citecolor = blue,
anchorcolor = blue]{hyperref}

\definecolor{lime}{HTML}{A6CE39}
\DeclareRobustCommand{\orcidicon}{%
	\begin{tikzpicture}
		\draw[lime, fill=lime] (0,0) 
		circle [radius=0.16] 
		node[white] {{\fontfamily{qag}\selectfont \tiny ID}};
		\draw[white, fill=white] (-0.0625,0.095) 
		circle [radius=0.007];
	\end{tikzpicture}
	\hspace{-2mm}
}

\foreach \x in {A, ..., Z}{%
	\expandafter\xdef\csname orcid\x\endcsname{\noexpand\href{https://orcid.org/\csname orcidauthor\x\endcsname}{\noexpand\orcidicon}}
}

\usepackage[T1]{fontenc}
\usepackage[latin9]{inputenc}
\usepackage{color}
\usepackage{array}
\usepackage{amstext}
\usepackage{graphicx}
\usepackage{esint}
\usepackage{rotating}
\usepackage{appendix}
\usepackage{float}
\usepackage{subcaption}
\usepackage{amsmath}

\usepackage[font=small,labelfont=bf]{caption}
\usepackage{xcolor}
\usepackage{ulem}

\makeatletter

\@ifundefined{textcolor}{}
{%
	\definecolor{BLACK}{gray}{0}
	\definecolor{WHITE}{gray}{1}
	\definecolor{RED}{rgb}{1,0,0}
	\definecolor{GREEN}{rgb}{0,1,0}
	\definecolor{BLUE}{rgb}{0,0,1}
	\definecolor{CYAN}{cmyk}{1,0,0,0}
	\definecolor{MAGENTA}{cmyk}{0,1,0,0}
	\definecolor{YELLOW}{cmyk}{0,0,1,0}
}

\@ifundefined{definecolor}
{\usepackage{color}}{}
\@ifundefined{definecolor}
{\usepackage{color}}{}
\makeatother
\usepackage{babel}

\newcommand{\beq}{\begin{equation}}
	\newcommand{\eeq}{\end{equation}}

\def\Re{{\cal R \mskip-4mu \lower.1ex \hbox{\it e}\,}}
\def\Im{{\cal I \mskip-5mu \lower.1ex \hbox{\it m}\,}}

\def\tev{\,{\ifmmode\mathrm {TeV}\else TeV\fi}}
\def\gev{\,{\ifmmode\mathrm {GeV}\else GeV\fi}}
\def\mev{\,{\ifmmode\mathrm {MeV}\else MeV\fi}}

\begin{document}

	\title{ Alleviating Cosmological Tensions with the Hadrosymmetric Twin Higgs }

	\author{Mohammad Soroori Sotudeh$^{1}$\orcidA{}}
	\email{mohammad.soroori@ut.ac.ir}

	\author{ Zahra Davari$^{2}$\orcidB{}}
	\email{zahradavari@kias.re.kr}

	\author{Sara Khatibi$^{1}$\orcidC{}}
	\email{sara.khatibi@ut.ac.ir}

	\author{Nima Khosravi$^{3}$\orcidD{}}
	\email{nima@sharif.edu}

	\affiliation {
		$^{(1)}$Department of Physics, University of Tehran, North Karegar Ave., Tehran 14395-547, Iran   \\ 
		$^{(2)}$School of Physics, Korea Institute for Advanced Study (KIAS), 85 Hoegiro, Dongdaemun-gu, Seoul, 02455, Korea   \\
		$^{(3)}$Department of Physics, Sharif University of Technology, Tehran 11155-9161, Iran
	}


	\begin{abstract}
		\label{abstract}
		The Hadrosymmetric Twin Higgs (HTH) model provides a natural solution to the little hierarchy problem by incorporating all three generations of quarks in a twin sector. Unlike other Twin Higgs scenarios, such as the Mirror Twin Higgs (MTH), the HTH framework avoids introducing additional light states or radiation and thus remains consistent with stringent bounds on the effective number of relativistic species, $\Delta N_{\rm eff}$. Its particle content and interactions also make it difficult to probe at colliders, highlighting the importance of cosmological tests. In this work, we study the cosmological implications of the HTH model, focusing on the persistent tensions in the Hubble constant ($H_0$) and the matter clustering amplitude ($\sigma_8$). Implementing the HTH sector in a Boltzmann code and confronting it with cosmic microwave background (CMB) data and local $H_0$ measurements, we find that HTH scenario partially reduces the Hubble tension from more than $4\sigma$ to about $2.5\sigma$, while also alleviating the $\sigma_8$ discrepancy. These results demonstrate that the HTH framework not only addresses naturalness in particle physics but also offers a viable route to mitigating current cosmological tensions, thereby strengthening the link between fundamental theory and precision cosmology.
	\end{abstract}  


	\maketitle


	\newpage
	\section{Introduction} \label{intro}

	The Standard Model (SM) of particle physics is one of the most successful and well-established theories in modern physics, 
	yet it leaves unresolved questions, most notably the hierarchy problem. 
	Numerous Beyond Standard Model (BSM) theories have been developed to address this issue, 
	many of which rely on symmetry-based approaches to protect the Higgs mass and predict the existence of new partners for the SM particles. 
	Despite extensive searches, the absence of experimental evidence for such particles at the Large Hadron Collider (LHC) 
	has shifted attention toward models containing new particles neutral under the SM gauge symmetries.
	Among these, Twin Higgs (TH) models present an appealing solution to the little hierarchy problem 
	by introducing a new twin sector, including particles uncharged under SM gauge interactions~\cite{Chacko:2005pe,Chacko:2005un}.

	The Higgs sector of the TH models has an approximate global symmetry wherein 
	the SM Higgs boson is a pseudo-Nambu-Goldstone boson (pNGB) of the spontaneous 
	symmetry breaking (SSB) of this global symmetry, which allows the Higgs boson to remain naturally light. 
	However, beyond 5 to 10 TeV, a UV-complete model is necessary because the approximate global 
	symmetry is not radiatively stable. Supersymmetry~\cite{Chang:2006ra,Craig:2013fga,Falkowski:2006qq,Badziak:2017syq,Badziak:2017kjk,Badziak:2017wxn} 
	and Composite Higgs models~\cite{Batra:2008jy,Barbieri:2015lqa,Low:2015nqa} 
	are examples of such UV completions. 
	In its most straightforward form, all the SM particles and symmetries are duplicated in the twin sector, 
	known as the Mirror Twin Higgs (MTH) model~\cite{Chacko:2005pe}. 
	This means that a discrete $\mathbb{Z}_2$ symmetry links the particles 
	and symmetries of the SM and TH sectors. Since the particles in the MTH sector carry no charge under SM symmetries, 
	they are difficult to produce at the LHC~\cite{Burdman:2014zta}.

	However, this model is affected by the bound on the effective SM neutrino degrees of freedom, $\Delta N_{\rm eff}$. 
	The expansion rate of the universe is increased by the existence of additional relativistic degrees of freedom 
	like twin photons and twin neutrinos. 
	However, in a minimal scenario,  to evade the little hierarchy problem, just the third generation of twin quarks, 
	charged under twin $SU(3)^{'}_{C}$ and twin $SU(2)^{'}_{L}$ gauge symmetries~\footnote{The twin particles, couplings, and symmetries are denoted by a prime.}
	and have almost the same SM coupling values, are required and 
	other features of the twin sectors are not constrained very hard~\cite{Craig:2014roa,Craig:2014aea}.
	So, some other varieties of the TH model have been proposed~\cite{Barbieri:2016zxn,Craig:2016lyx,Chacko:2016hvu,Badziak:2025fdp},
	like the fraternal twin Higgs~\cite{Craig:2015pha}, the vector-like twin Higgs~\cite{Craig:2016kue} 
	and the Hadrosymmetric Twin Higgs~\cite{Freytsis:2016dgf}.

	In contrast to the MTH model, in the Hadrosymmetric Twin Higgs (HTH) model, discrete $\mathbb{Z}_2$ symmetry is hardly broken, 
	so the twin sector is not simply a perfect copy of the SM spectrum, and it just contains a mirror copy of the SM hadrons~\cite{Freytsis:2016dgf}.
	The twin sector includes all three generations of quarks; however, it does not contain any light states or radiation.
	Furthermore, the twin hypercharge is considered to be a global symmetry instead of a gauge symmetry, 
	which means that the twin photon does not appear in the twin spectrum. As a result, 
	the twin $SU(2)^{'}_{L}$ gauge bosons possess equal masses.

	Since the HTH model does not introduce any additional light particles, it avoids current bounds on $\Delta N_{\rm eff}$. 
	At the same time, its direct collider signatures are extremely elusive~\cite{Schwaller:2015gea,Cohen:2015toa}, 
	making it one of the least experimentally accessible realizations among TH scenarios. 
	This motivates exploring its implications in cosmology, where precise measurements of the expansion history 
	and structure growth offer a complementary probe.

	In particular, modern cosmology currently faces two persistent anomalies within the $\Lambda$CDM framework.
	The first is the Hubble tension, a $\sim 4\sigma$ discrepancy between early-time (CMB) and 
	late-time (distance-ladder) determinations of $H_{0}$~\cite{Riess:2019cxk,Planck:2018vyg}. 
	The second is the $\sigma_{8}$ tension, a $\sim 2$--$3\sigma$ mismatch between the amplitude 
	of matter clustering inferred from large-scale structure surveys and that predicted by the CMB~\cite{Heymans:2020gsg,DES:2021wwk}. 
	If not attributable to systematics, these discrepancies may signal physics beyond the standard model of cosmology and or particle physics.  
	There are numerous studies in this area, including modifications to 
	dark energy~\cite{Banihashemi:2018has, Vagnozzi:2021gjh}, 
	dark matter~\cite{Buen-Abad:2015ova, Bagherian:2024obh, Davari:2022uwd, Arabameri:2023who, Ashoorioon:2023jwf}, 
	and even explorations of the physics of the early universe~\cite{Kamionkowski:2022pkx,  Poulin:2023lkg, Jedamzik:2020krr} 
	(See Ref.~\cite{CosmoVerseNetwork:2025alb} for a comprehensive review). 
	
	Cosmological extensions of the TH framework are 
	therefore particularly interesting, as recent work has shown that the
	MTH scenario can simultaneously alleviate both tensions~\cite{Bansal:2021dfh}.
	The authors carried out detailed numerical analyses of twin sector processes, 
	including twin Big Bang nucleosynthesis, twin recombination, and the evolution of perturbations. 
	By fitting the model to CMB and large scale structure data, 
	they showed that the MTH framework not only remains consistent with current cosmological constraints 
	but can also simultaneously alleviate the Hubble and $\sigma_8$ tensions.

	In this paper, we investigate the cosmological phenomenology of the HTH model. 
	We implement the HTH sector into the publicly available numerical 
	code \texttt{CLASS}\footnote{\label{myfootnote0}\url{https://github.com/lesgourg/class_public}}(the Cosmic Linear Anisotropy Solving System)~\citep{Lesgourgues:2011rh}
	and to perform a Monte Carlo Markov Chain (MCMC) analysis with a Metropolis-Hasting algorithm using
	the code \texttt{MONTEPYTHON-v3}\footnote{\label{myfootnote}\url{https://github.com/baudren/montepython_public}}~\citep{Audren:2012wb,Brinckmann:2018cvx}  using the Planck 2018 high-$\ell$ CMB TT, TE, EE + low-$\ell$ TT, EE + lensing data~\citep{Aghanim:2018eyx}. 
	Our analysis shows that the HTH framework, originally motivated by particle physics considerations, 
	can also reduce both the Hubble and $\sigma_8$ tensions, thereby connecting the resolution of 
	the little hierarchy problem to present-day cosmological observations.

	The structure of the paper is as follows. Section~\ref{sec:model} describes the details of the HTH model. 
	Section~\ref{sec:Cosmology} presents its cosmological phenomenology. 
	Finally, Section~\ref{sec:summary} summarizes our conclusions.

	\section{The Model}\label{sec:model}

	In this section, we provide a detailed explanation of the HTH model, as introduced in the paper~\cite{Freytsis:2016dgf}.
	The Higgs potential of the model demonstrates an approximate $SU(4)$ symmetry. 
	The SM Higgs boson, denoted as $H$, and the twin Higgs boson, denoted as $H'$, 
	exist within the fundamental representation of the $SU(4)$ symmetry, $(H, H')$. 
	Following SSB in both sectors,\footnote{The vacuum expectation values (vev) for the SM and twin sectors are represented respectively as: 
		$v \equiv\langle H\rangle,$ and $v' \equiv\langle H'\rangle$.} 
	two physical degrees of freedom remain: a pNGB and a radial mode. 
	The pNGB is naturally light and corresponds to the observed boson at the LHC. 
	The mixing between the SM Higgs and the twin Higgs serves as the portal between the two sectors. 
	Consequently, there will be a deviation in the Higgs couplings; to avoid significant discrepancies 
	from the current experimental Higgs branching ratios, $v'/v$ must exceed 3~\cite{Craig:2015pha,Burdman:2014zta}.

	Additionally, like other twin Higgs models, the HTH model features an approximate $\mathbb{Z}_2$ symmetry between 
	the Standard Model and twin sectors, which helps to cancel quadratic divergences. 
	To achieve this cancellation, the top quark Yukawa, 
	weak, and color gauge couplings in two sectors should satisfy the following relations respectively~\cite{Craig:2015pha,Freytsis:2016dgf},
	\beq
	\frac{\left|y_t-y_t^{\prime}\right|}{y_t} \sim 0.01, \quad \frac{\left|g_2-g_2^{\prime}\right|}{g_2} \sim 0.1, \quad \text { and } \quad \frac{\left|g_3-g_3^{\prime}\right|}{g_3} \sim 0.1,
	\label{Eq:limit}
	\eeq
	while in the above relations, the cut-off of the twin Higgs sector is set to be around 5 TeV.
	There are no restrictions on the other Yukawa couplings and hypercharge coupling.
	Unlike other twin Higgs models, the HTH scenario includes all three generations of quarks 
	but does not feature any light lepton generations within the twin spectrum.

	The model's gauge symmetry is expressed as $SU(3)^{'}_{C} \times SU(2)^{'}_{L}$, 
	while its twin hypercharge symmetry is a global symmetry rather than gauge symmetry, 
	resulting in all twin weak gauge bosons have similar masses,
	$m_{W' , Z'} = \frac{1}{2} g'_2  v^{'}$.
	So according to this fact and Eq.~\ref{Eq:limit}, it is assumed that all particles, gauge bosons and quarks, in twin spectrum are heavier 
	than their SM counterparts with a factor $v'/v$.

	The dark hadron spectrum in the HTH model resembles its SM counterpart. 
	According to Eq.~\ref{Eq:limit}, due to a slight deviation in the twin QCD coupling 
	(approximately $10\%$), the twin QCD confinement scale changes only slightly. 
	Based on the two-loop renormalization group running of the strong coupling, 
	the ratio of the confinement scale ($\Lambda'_{\rm{QCD}}/\Lambda_{\rm{QCD}}$) varies between 0.2 and 5~\cite{Freytsis:2016dgf,Bansal:2021dfh}.

	The classification of light and heavy twin quarks is determined by comparing their masses to the twin QCD confinement scale. 
	In the HTH scenario, three light twin quarks are considered, leading to the formation of twin pions within the dark sector. 
	These twin pions, a twin isospin triplet \((\pi'^{\pm}, \pi'^{0})\), 
	are the lightest pNGBs and can have masses well below the twin confinement scale.
	However, to remain consistent with constraints on the \(\Delta N_{\rm eff}\), 
	the twin pion mass must exceed the temperature of Big Bang Nucleosynthesis (BBN), \(T_{\rm BBN} \sim \mathcal{O}(\mathrm{MeV})\).

	Since the charged twin pions \(\pi'^{\pm}\) carry a conserved global \(U(1)\) charge, 
	they are stable; however, their relic abundance is expected to be negligible~\cite{Freytsis:2016dgf}. 
	The twin sector also includes other light and heavy hadronic states, likewise the SM spectrum,
	such as the twin proton and twin neutron. The twin proton is stable due to baryon number conservation, 
	while the twin neutron remains stable in the absence of light twin-sector leptons.
	Heavier twin hadrons eventually decay into the stable constituents of the dark sector: twin protons, twin neutrons, and twin pions.

	Furthermore, to avoid cosmological overclosure or an early matter-dominated era, 
	the neutral twin pion, \(\pi'^0\), must decay efficiently into SM degrees of freedom before BBN, 
	requiring a lifetime \(\tau_{\pi'^0} < \tau_{\rm BBN} \sim 1\,\mathrm{s}\). 
	As a pNGB and pseudoscalar, \(\pi'^0\) cannot efficiently decay to the SM states via the Higgs portal, 
	since such a decay necessitates both parity and twin isospin violation, leading to strong suppression. 
	To ensure prompt decay before BBN, an additional portal between the twin and SM sectors is required 
	to allow the \(\pi'^0\) to decay sufficiently rapidly.

	In the mass range \( m_{\pi'^0} < 3m_{\pi^0} \), it is necessary to have a UV completion twin-SM portal below the TeV scale,
	potentially accessible at the LHC. Since the decay requires twin-isospin violation, 
	it is most efficient when the twin pion mixes with the lightest SM pseudoscalar, the \(\pi^0\). 
	Furthermore, the dominant contribution to the twin pion to SM decay amplitude arises from an off-shell \(\pi^0\),
	$\pi'^0 \rightarrow \pi^{0*} \rightarrow \rm{SM}$.  
	As estimated in Ref.~\cite{Freytsis:2016dgf}, for \( m_{\pi'^0} < 3m_{\pi^0} \) (\( \sim 400\,\mathrm{MeV} \)), 
	the diphoton mode is expected to be the dominant decay channel for the twin pion.

	In the HTH framework, several candidates for dark matter are theoretically viable. 
	One primary scenario is the twin WIMP (T-WIMP), a symmetric thermal relic that freezes out via twin electroweak interactions. 
	In this model, T-WIMP annihilation produces a `dark shower' of twin-sector quarks that hadronize into stable twin nucleons and twin pions, 
	the latter of which decay into SM photons through various portals~\cite{Freytsis:2016dgf}. 
	Alternatively, if a twin baryon asymmetry is present in the early universe, 
	the stable twin proton and twin neutron can serve as twin baryonic asymmetric dark matter (ADM)~\cite{GarciaGarcia:2015pnn,Farina:2015uea}.

	In the following section, we examine the cosmological implications of the HTH model in light of the features discussed above.
	We first describe how the HTH framework is implemented in the Boltzmann solver \texttt{CLASS},
	with particular attention to the decays of twin neutral pions, 
	which govern energy transfer between the twin and visible sectors and impact both the expansion history and structure formation. While the  stable species, mentioned above, play the role of  cold dark matter in the \texttt{CLASS}.  
	In contrast to the MTH scenario studied in Ref~\cite{Bansal:2021dfh}, where light twin states dominate the cosmology, 
	the HTH model contains no such states, making twin hadron dynamics the key driver of its cosmological effects. 
	We then assess whether these features can help to alleviate current cosmological anomalies, the $H_0$ and $\sigma_8$ tensions.

	\section{Cosmological phenomenology of HTH}\label{sec:Cosmology}

	As discussed in the introduction, the standard $\Lambda$CDM model cannot  
	address the Hubble and $\sigma_8$ tensions. This motivates the investigation of 
	BSM scenarios where new degrees of freedom can affect cosmological 
	observables. The HTH model, while originally motivated 
	by naturalness in particle physics, provides a natural testing ground in this context.

	To assess its cosmological impact, we implement the HTH framework in the Boltzmann 
	solver \texttt{CLASS}~\citep{Lesgourgues:2011rh}. For simplicity, and following the treatment 
	of Ref.~\citep{Bansal:2021dfh} in the case of the MTH, we model 
	the twin-sector states (twin neutral pions) as an effective decaying dark matter (DDM) component, while 
	accounting for possible interactions that allow the dark sector to produce visible 
	photons. This approach captures the leading phenomenological effects without relying 
	on the detailed microphysics of the dark hadron spectrum. In background of the \texttt{CLASS} code, 
	we used relations describing DDM and its contribution to extra radiation ($\rm \bar{R}$) given
	\begin{eqnarray}\label{rhodm}
		&&\dot{\rho}_{_{\rm DDM}}+3H\rho_{_{\rm DDM}}=-\gamma_{_{\rm DDM}}\rho_{_{\rm DDM }},\nonumber\\
		&&\dot{\rho}_{_{\rm \bar{R}}}+4H\rho_{_{\rm \bar{R}}}=\gamma_{_{\rm DDM}}\rho_{_{\rm DDM }},
	\end{eqnarray}
	where $\gamma_{_{\rm DDM}}$ is the decay rate and $\rho_{_{\rm \bar{R} }}$ represents the produced visible photons by the twin sector. 
	When the decay rate is non-zero ($\gamma_{_{\rm DDM}} \neq 0$),  these
	produced  photons effectively add to the photon radiation content of the universe and enters 
	the cosmological background evolution. As a result, the parameters associated with the decay rate 
	directly affect both the expansion history and the effective radiation fraction of the universe.

	Hence, the Boltzmann equations describing the evolution of photon perturbations in HTH model are modified accordingly as,
	\begin{eqnarray}
		\dot{\delta}_{\gamma} &=& -\tfrac{4}{3}\theta_{\gamma} + 4\dot{\phi} +
		a\gamma_{\rm DDM} \frac{\rho_{\rm DDM}}{\rho_{\gamma}}
		\left(\delta_{\rm DDM} - \delta_{\gamma} + \psi\right), \nonumber\\
		\dot{\theta}_{\gamma} &=& k^2\!\left(\tfrac{\delta_{\gamma}}{4} - \sigma_{\gamma} + \psi\right)
		+ a n_e \sigma_T \left(\theta_b - \theta_\gamma\right)
		- a\gamma_{\rm DDM}\frac{3\rho_{\rm DDM}}{\rho_\gamma}
		\left(\tfrac{4}{3}\theta_{\gamma} - \theta_{\rm DDM}\right), \nonumber\\
		\dot{\sigma}_{\gamma} &=& \tfrac{4}{15}\!\left(\theta_{\gamma} - \tfrac{9}{8}kF_{\gamma 3}\right)
		- \tfrac{9}{10} a n_e \sigma_T \sigma_{\gamma}
		+ \tfrac{1}{20} a n_e \sigma_T \left(G_{\gamma 0}+G_{\gamma 2}\right)
		- a\gamma_{\rm DDM}\frac{\rho_{\rm DDM}}{\rho_{\gamma}}\sigma_{\gamma}, \nonumber\\
		\dot{F}_{\gamma l} &=& \tfrac{k}{2l+1}\!\left[l F_{\gamma(l-1)} - (l+1)F_{\gamma(l+1)}\right]
		- a n_e \sigma_T F_{\gamma l}, \qquad l \geq 3, \nonumber\\
		\dot{G}_{\gamma l} &=& \tfrac{k}{2l+1}\!\left[l G_{\gamma(l-1)} - (l+1)G_{\gamma(l+1)}\right] 
		+ a n_e\sigma_T \Bigg[-G_{\gamma l}
		+ \tfrac{1}{2}\left(F_{\gamma 2}+G_{\gamma 0}+G_{\gamma 2}\right)
		\left(\delta_{l0}+\tfrac{\delta_{l2}}{5}\right)\Bigg].
	\end{eqnarray}	

	\begin{figure}
		\begin{center}
			\includegraphics[scale=1]{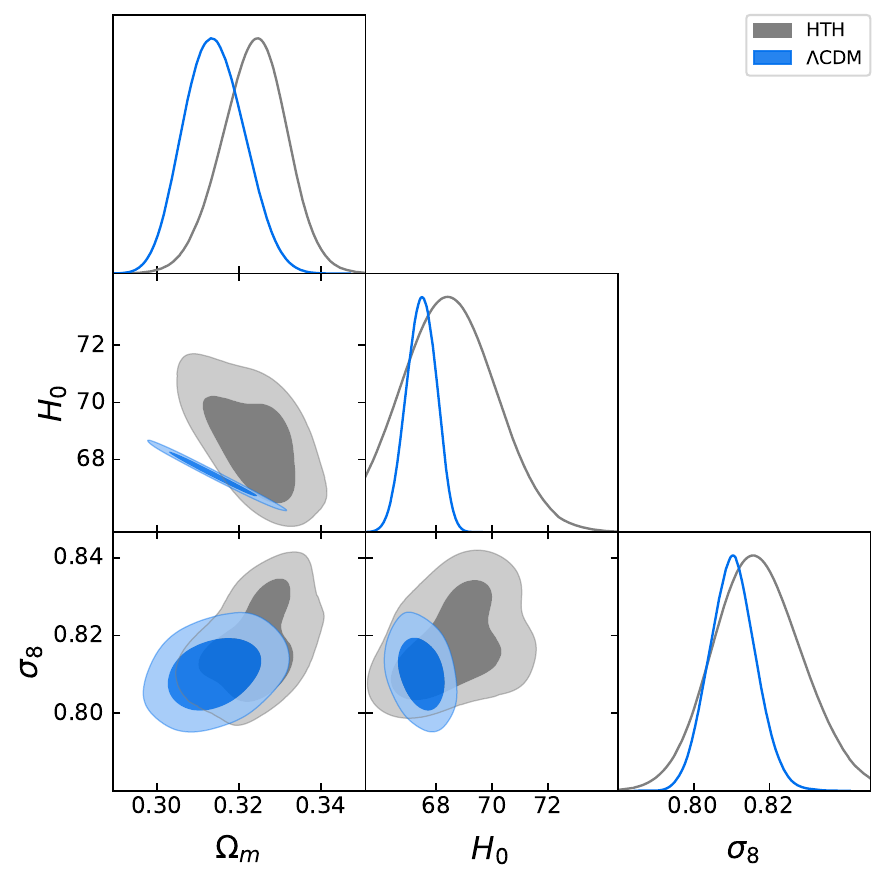}   
			\caption{
				Likelihood contours for the $\Lambda$CDM (blue) and HTH (grey) models obtained using only the ``only CMB'' dataset. 
				The HTH contours are broader due to the additional free parameters. 
				Compared to $\Lambda$CDM, the Hubble tension is reduced, and the $\sigma_8$ discrepancy is milder. 
			}   
			\label{fig:HTH}
		\end{center}
	\end{figure}

	\begin{figure}
		\begin{center}
			\includegraphics[scale=1]{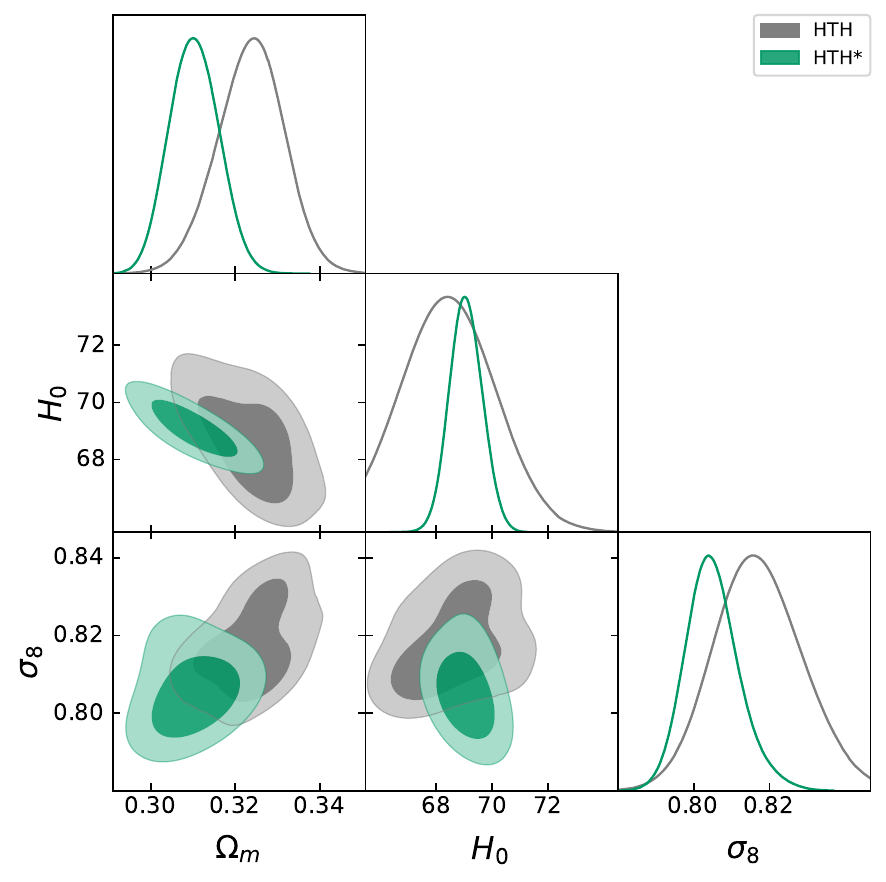}   
			\caption{ 
				Likelihood contours for the HTH model are presented. The posteriors are deduced by implementing ``only CMB'' and ``CMB+$H_0$'' datasets labeled as HTH and HTH*, respectively. It is obvious that the HTH* shows a tighter likelihoods as it is expected. 
			}   
			\label{fig:HTH2}
		\end{center}
	\end{figure}

	These equations are expressed in Conformal Newtonian gauge. Here, $\delta_{\gamma}$, $\theta_{\gamma}$, and $\sigma_{\gamma}$ 
	denote the density perturbation, velocity divergence, and shear stress of photons, respectively. $F_\gamma$ is the $\ell$-th moment 
	of photon temperature perturbations and $G_\gamma$ is the difference of the two linear polarization components. 
	$k$ is the comoving wavenumber, $n_e$ is the number density of  electrons, 
	and $\sigma_T$ is the Thomson scattering cross-section for the twin sector. 
	An overhead dot denotes derivatives with respect to conformal time.

	\begin{figure}
		\begin{center}
			\includegraphics[scale=0.8]{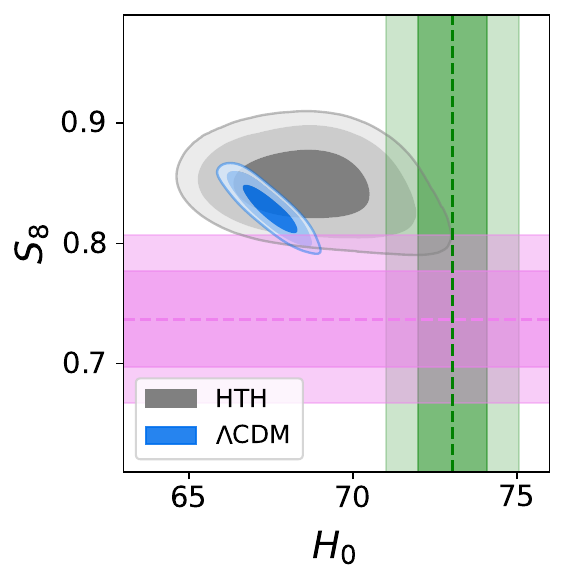}   
			\includegraphics[scale=0.8]{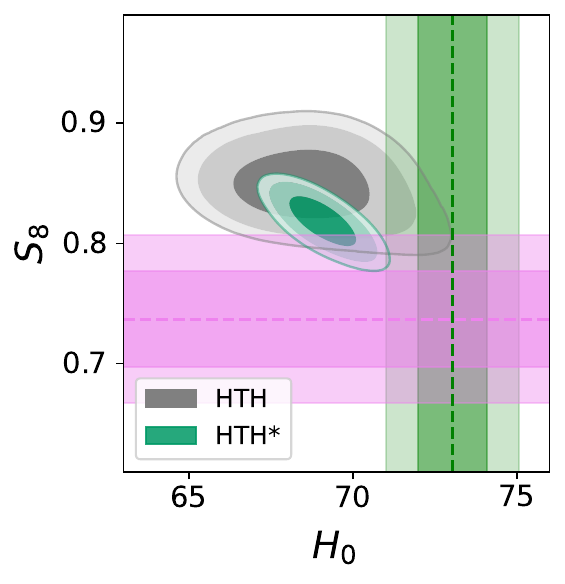}

			\caption{
				In the left panel: preferred ranges of $S_8 = \sigma_8\sqrt{\Omega_m/0.3}$ and $H_0$ for the HTH and $\Lambda$CDM models using the Planck 2018 dataset. 
				For comparison, the SH0ES measurement of $H_0$ (green) and the KV450 measurement of $S_8 = 0.737^{+0.040}_{-0.036}$ (violet) 
				are also shown \citep{Hildebrandt:2018yau}. Grey and red contours correspond to HTH and $\Lambda$CDM, respectively. In the right panel, the HTH and HTH* contours are plotted in grey and green, respectively. The ``only CMB'' data set is used to constrained $\Lambda$CDM and HTH models while the ``CMB+$H_0$'' dataset is used to constrained HTH*.}
			\label{HTHS8}
		\end{center}
	\end{figure}

	\subsection{The HTH model and the Hubble tension: pre numerical results}\label{pre-res}
	
		Before presenting the numerical results, it is informative to discuss on how the HTH model may reduce the  the Hubble tension. 
		As we already mentioned, the Hubble tension refers to the discrepancy between the locally measured value of the present-day Hubble parameter, $H_0$, 
		and the value inferred from the CMB within the framework of the $\Lambda$CDM model.
		This discrepancy may originate from modifications of either early-universe physics, 
		late-universe physics, or a combination of both~\cite{CosmoVerseNetwork:2025alb}.  A key point is that CMB observations do not directly constrain $H_0$. 
		Instead, they precisely measure the angular scale of the acoustic peaks, $\theta_\ast$, which corresponds to the angular 
		size of temperature fluctuations at recombination. This observable depends on both pre- and post-recombination physics. 
		More specifically, $\theta_\ast$ is determined by the ratio of two distances:
		\begin{itemize}
			\item the comoving sound horizon at recombination, $r_s$, and
			\item the comoving angular diameter distance to the last-scattering surface.
		\end{itemize}
		In our case, the HTH model modifies the early-time cosmological dynamics, 
		and therefore our focus is on the sound horizon, $r_s$, which encodes the physics from the Big Bang up to photon decoupling. It is given by
		\begin{equation}
			r_s = \int_{z_\ast}^{\infty} \frac{c_s(z)}{H(z)} \, dz ,
		\end{equation}
		where $c_s(z)$ is the sound speed in the photon--baryon plasma and $H(z)$ is the Hubble expansion rate.
		A larger inferred value of $H_0$ from CMB data requires a smaller sound horizon $r_s$. Therefore, any model that modifies either the sound speed $c_s(z)$ 
		or the expansion history $H(z)$ prior to recombination can potentially alleviate the Hubble tension.
		In the HTH framework, both effects are present. The sound speed $c_s$ is modified due to a non-trivial photon source arising from the twin sector. We expect both effects to contribute to reducing $r_s$. Since the late time cosmology is fixed by the data then we  assume the cosmological parameters (specifically the fractional densities $\Omega_i$'s) are comparable to the standard $\Lambda$CDM's ones. Now if we go backward in time, both models are same till the redshift that the decaying dark matter pumps the photon density. Before that redshift, the ratio $R=3\rho_b/4\rho_\gamma$  is larger in the HTH in comparison to the $\Lambda$CDM since the photons transform to the (decaying) dark matter particles (when we go backward in time). This means a less $r_s$ for the HTH model in comparison to the $\Lambda$CDM and consequently a higher value of $H_0$.
		In addition, the expansion history is altered because of an additional matter component that later dilutes into radiation, 
		thereby changing the pre-recombination evolution of $H(z)$. This effect is in the direction of larger $H(z)$'s for the HTH in comparison to $\Lambda$CDM. Again, assume both models are sharing (almost) the same fractional densities. Then $H^2_{HTH}(z) \approx H^2_{\Lambda CDM}(z)+\Omega_{_{\rm DDM}}(z)H^2_{\Lambda CDM}(z)$ which reduces to $(H_{HTH}-H_{\Lambda CDM})/H_{\Lambda CDM}\approx \Omega_{_{\rm DDM}}/2$. Note that we assumed $\Omega_{_{\rm DDM}}\ll 1$ which is what we expect. We note that the above reasoning is based on simplified analytical arguments. The real physics behind the interactions and their effects on the CMB can be very complicated and the final answer should be considered after the numerical results, presented below.


	\subsection{Numerical Results}
	
	We check the HTH model against the CMB dataset (Planck 2018 high-$\ell$ CMB TT, TE, EE + low-$\ell$ TT, EE + lensing), firstly. 
	This can show us if the HTH model has the potential to solve the $H_0$ and $\sigma_8$ tensions or not. 
	We use Bayes's theorem to compute the posterior distributions of the model parameters. 
	Beyond the six standard $\Lambda$CDM parameters, $(\Omega_b,\Omega_{DM},100\theta_{\rm MC},\ln 10^{10}A_s,n_s,\tau_{\rm reio})$, 
	the HTH model introduces $(\Omega_{\rm DDM},\gamma_{_{\rm DDM}})$, with flat priors $\Omega_{\rm DDM}\in[0,0.1]$ 
	and $\gamma_{_{\rm DDM}}\times 10^{-8}\in[0,2\times10^3]$. We consider the decay constant of $\gamma_{_{\rm DDM}}$ is $\rm kms^{-1}Mpc^{-1}$, 
	the same unit as $H_0$ in \texttt{CLASS}. Convergence of the MCMC chains is verified via 
	the Gelman-Rubin criterion ($R-1<0.01$ for all parameters), with an average acceptance rate of $\sim0.2$.

	Figure~\ref{fig:HTH} shows the posterior contours for $\sigma_8$, $\Omega_M$ and $H_0$ for the HTH model and the $\Lambda$CDM for comparison. Note that these posteriors are deduced by just implementing the CMB dataset. Even in the absence of additional relativistic species, the HTH framework leads to a statistical relaxation of the Hubble tension, reducing its significance from more than $4\sigma$ to approximately $2.5\sigma$. This effect arises mainly from a broadening of the posterior distributions, rather than a large shift in their mean values. This improvement is non-trivial: the model operates under tight CMB photon constraints, leaving minimal freedom for tuning. 
	Thus, the HTH provides a minimal yet impactful modification to the standard cosmology. It is instructive to clear our comparison between two models by reporting the chi-square values evaluated at the best-fit points. The $\chi^2_{\rm best}$ for $\Lambda$CDM and HTH models are $1389.47$ and $1385.18$ respectively. This means the HTH model works better than the $\Lambda$CDM. However, we have to emphasize that the HTH model has more free parameters and the $\chi^2_{\rm best}$ analysis is not sufficient. Another gauge which consider the number of model's free parameters is known as the $AIC$. The relative $AIC$ for the HTH and $\Lambda$CDM is $\Delta AIC(\text{HTH}-\Lambda\text{CDM})=0.29$ which means  the HTH model is as good as the $\Lambda$CDM. But it is crucial to remember (as stated above) that the HTH model allows for the higher $H_0$ values while the $\Lambda$CDM does not. This means we can check the HTH model against both CMB and local $H_0$ datasets while we are not allowed to do that for the $\Lambda$CDM.  In Figure \ref{fig:HTH2} we have shown the posteriors for the HTH* which is constrained simultaneously  by both the CMB and local $H_0$ datasets. Note that the models are same for the HTH and HTH* but the datasets which are used to deduce the parameters are different.

	\begin{table}
		\centering
		\caption{Cosmological parameters of the $\Lambda$CDM and HTH models inferred from the Planck 2018 dataset, and of the HTH* model inferred from both the Planck 2018 dataset and the local $H_0$ measurement. 
			Quoted uncertainties represent the $68\%$ (first) and $95\%$ (second) credible intervals. }
		\begin{tabular}{|l |c|c|c|c|}
			\hline
			& $\Lambda$CDM&HTH& HTH* \\
			\hline
			{\boldmath$\Omega{}_{cdm }$} & $0.258\pm 0.007\pm0.013$&$0.271\pm0.008\pm0.011$ & $0.262^{+0.007+0.019}_{-0.010-0.016}$\\
			\hline
			{\boldmath$\Omega{}_{b }  $} & $0.04839\pm 0.0006\pm0.0011$&$0.0514\pm0.0012\pm0.0022$ & $0.0507^{+0.0005+0.0031}_{-0.001-0.002}$ \\
			\hline
			{\boldmath$100\theta{}_{s }$} & $1.0419\pm 0.0003\pm0.0006        $&$1.0439\pm0.0010\pm0.0022$ & $1.0436^{+0.0005+0.0011}_{-0.0004-0.0011}$\\
			\hline    
			
			{\boldmath$ln(10^{10}A_{s })$} & $3.042\pm 0.014 \pm0.029$&$3.044\pm0.019\pm0.036  $& $3.030^{+0.028+0.046}_{-0.010-0.084}   $ \\
			\hline
			{\boldmath$n_{s }         $} & $0.966\pm 0.0040\pm0.0082$& $0.9595^{+0.0059}_{-0.0066}\pm0.012$& $0.957\pm 0.007 \pm 0.016         $\\
			\hline
			{\boldmath$\tau{}_{reio } $} & $0.0537\pm 0.0073\pm0.015$&$0.056\pm0.009\pm0.023   $ & $0.052^{+0.016+0.025}_{-0.005-0.047} $\\
			\hline
			{\boldmath$ \Omega{}_{\rm DDM } \times10^{+5}$} &-& $0.48\pm 0.25\pm0.61$&$0.53^{+0.09+0.54}_{-0.24-0.44}$\\
			\hline
			{\boldmath$\gamma_{\rm DDM } \times 10^{-8}$}&- & $1.68\pm 0.19\pm0.42              $ & $2.39^{+0.45+0.85}_{-0.39-0.91}    $\\
			\hline
			{\boldmath$\Omega{}_{m }  $} & $0.3067\pm 0.0071\pm0.014$& $0.3236\pm 0.0070\pm0.015       $& $0.314^{+0.008+0.022}_{-0.011-0.018} $\\
			\hline
			{\boldmath$H_0             $} & $68.1\pm 0.5\pm 1.1             $&$68.3\pm 1.0\pm2.6 $ &  $69.12^{+0.60+1.49}_{-0.83-2.66}     $\\
			\hline
			{\boldmath$\sigma_8         $} & $0.8225\pm 0.0062\pm0.012 $& $0.8163\pm0.012\pm0.020$&  $0.80\pm0.01 \pm0.02 $ \\
			\hline
		\end{tabular}\label{tab-best}
	\end{table}
	\begin{figure}
		\begin{center}
			\includegraphics[scale=0.3]{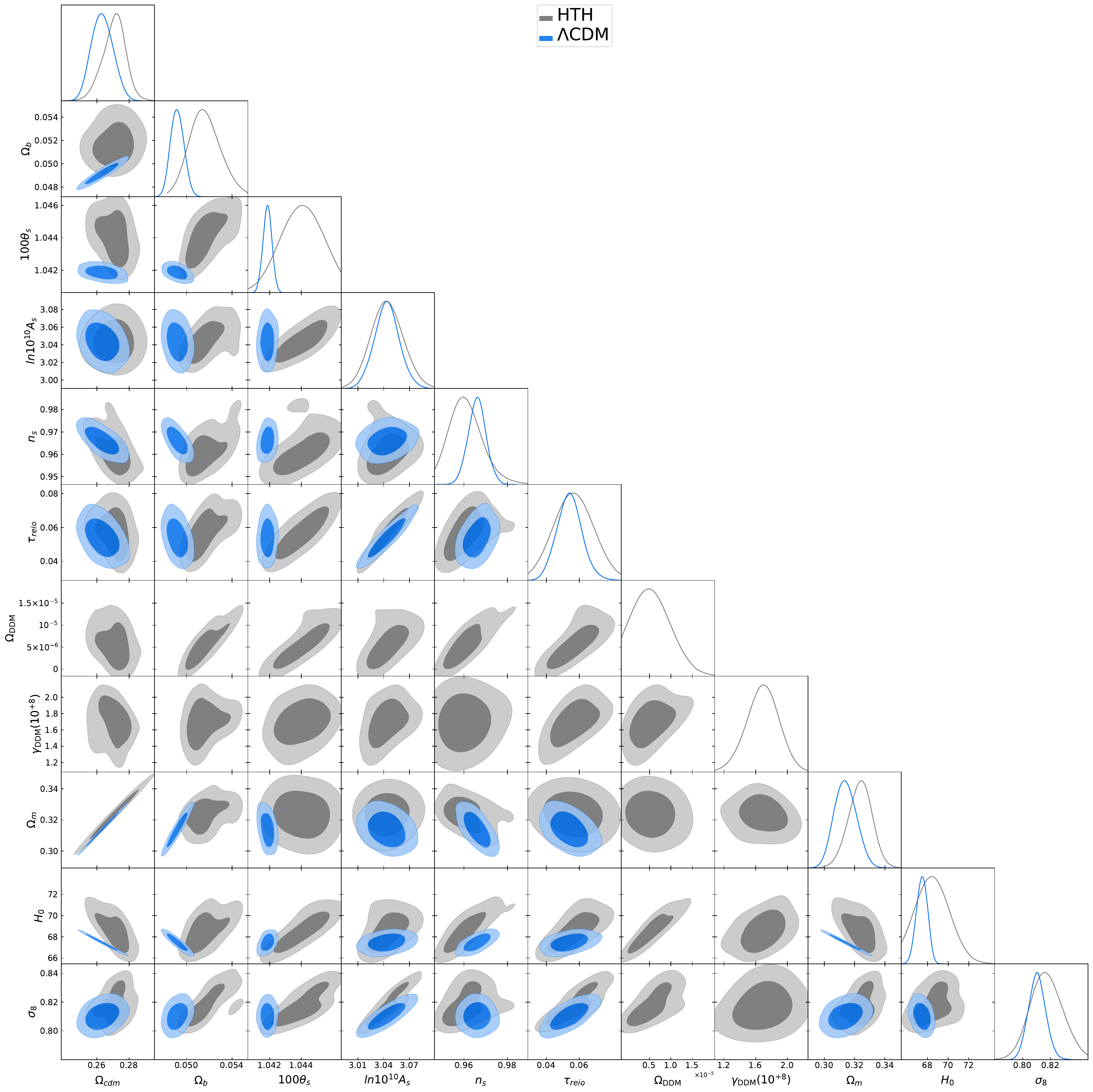}   
			\caption{Posterior distributions for all free parameters of the $\Lambda$CDM (blue) and HTH (grey) models using the ``only CMB'' dataset. }
			\label{fig:gHTH}
		\end{center}
	\end{figure}

	\begin{figure}
		\begin{center}
			\includegraphics[scale=0.3]{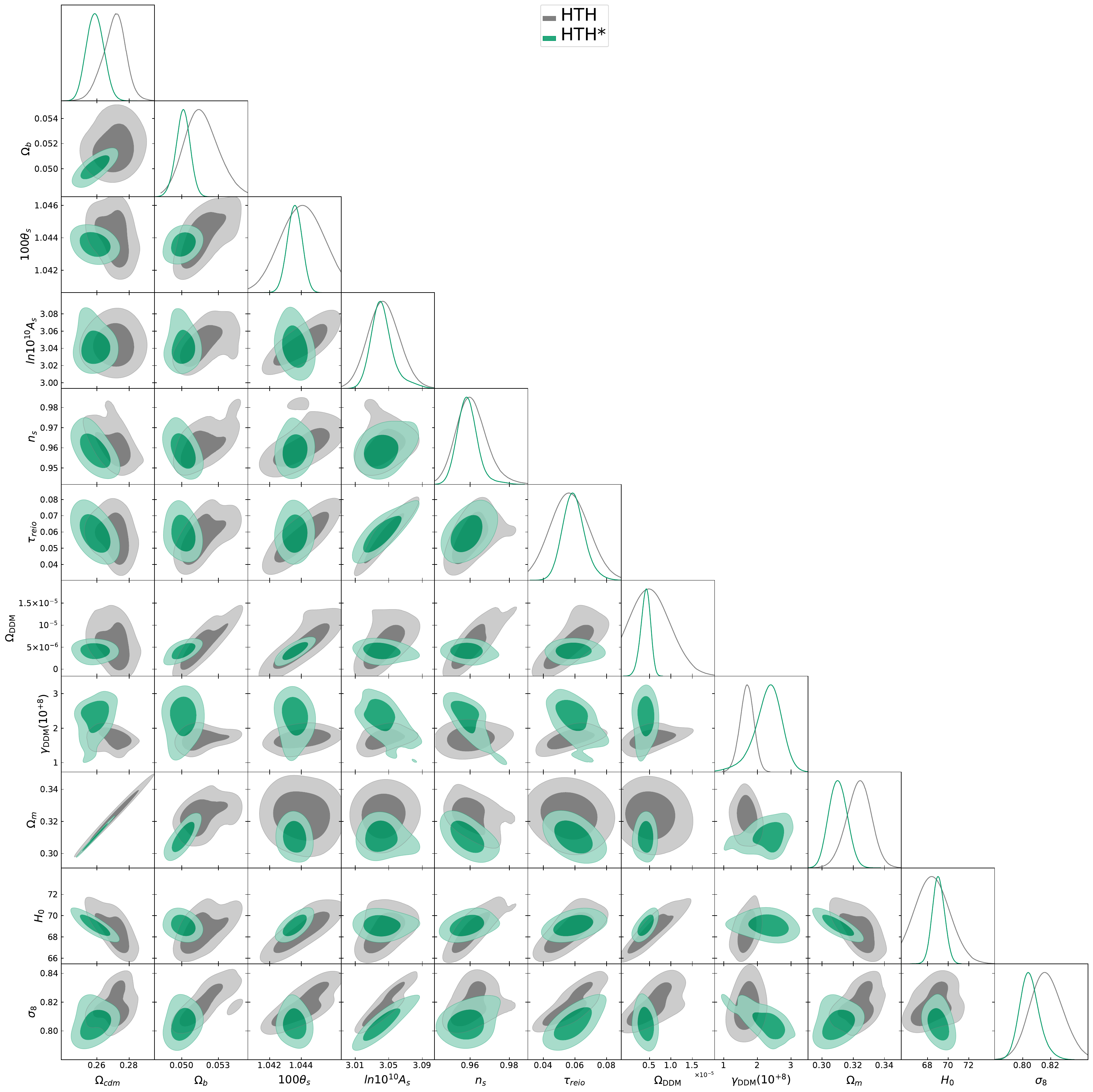}   
			\caption{Posterior distributions for all free parameters of the HTH (grey) and HTH* (green) variants using the ``only CMB'' and ``CMB+$H_0$'' datasets, respectively }
			\label{fig:gHTH2}
		\end{center}
	\end{figure}

	Incorporating the local Hubble measurement, $H_0 = 73.04 \pm 1.04,\mathrm{km,s^{-1},Mpc^{-1}}$~\citep{Riess:2021jrx}, 
	yields the HTH* scenario (Figure~\ref{fig:HTH2}), which does not favor lower $H_0$ values. 
	Unlike the MTH model, HTH introduces no extra ultra-relativistic degrees of freedom ($\Delta N_{\rm eff} \simeq 0$), 
	a feature that normally limits a model's ability to resolve the Hubble tension. 
	The fact that HTH nonetheless alleviates the tension highlights its genuinely non-trivial cosmological impact. 
	This result can be represented more clear if instead of $\sigma_8$ we employ $S_8=\sigma_8\sqrt{\Omega_M/0.3}$ 
	parametrization as shown in Figure~\ref{HTHS8}.

	Figures~\ref{fig:gHTH} and \ref{fig:gHTH2} summarize the posterior mean values and $1$-$2\sigma$ uncertainties for the standard $\Lambda$CDM model 
	and the HTH variants: HTH and HTH*. Compared to $\Lambda$CDM, the HTH models show small but significant shifts in $\Omega_b$ 
	and $\Omega_{\rm cdm}$, reflecting the impact of the decaying dark matter sector. Similarly, $100\theta_s$ 
	and $n_s$ exhibit slight adjustments to accommodate the modified expansion history and perturbation evolution in HTH, 
	while $\tau_{\rm reio}$ and $\ln 10^{10} A_s$ remain broadly consistent with standard predictions. 
	Overall, this comparison demonstrates that the HTH models remain fully compatible with current observational constraints, 
	while providing a modest but meaningful reduction in the Hubble tension.

	In Table~\ref{tab-best}, the mean values and the corresponding error bars for the parameters of $\Lambda$CDM, HTH and HTH* 
	are reported.
	
	\subsection{The HTH model and the Hubble tension: post numerical results}
	
\begin{figure}
		\begin{center}
			\includegraphics[scale=0.3]{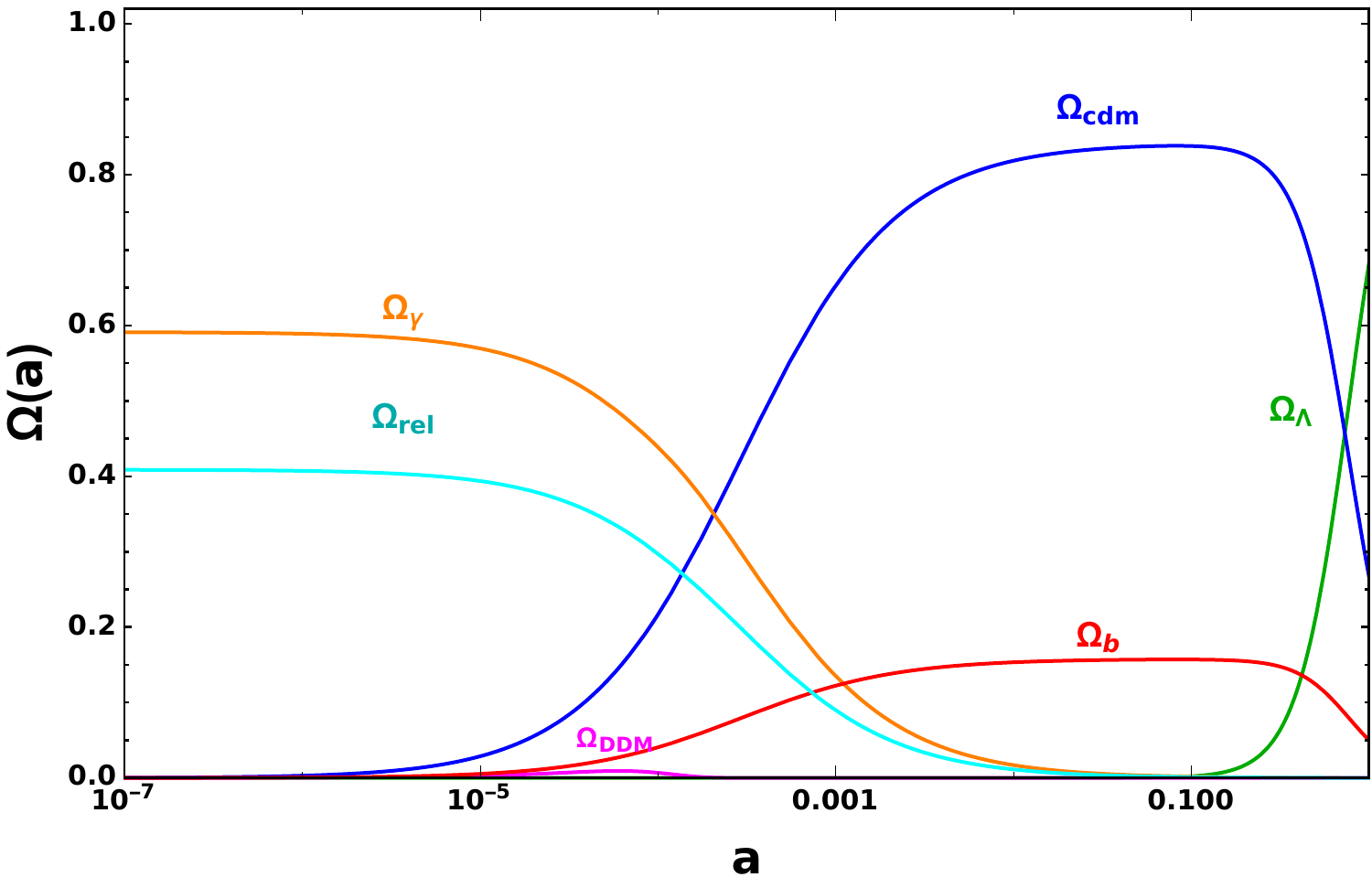}   
			\includegraphics[scale=0.3]{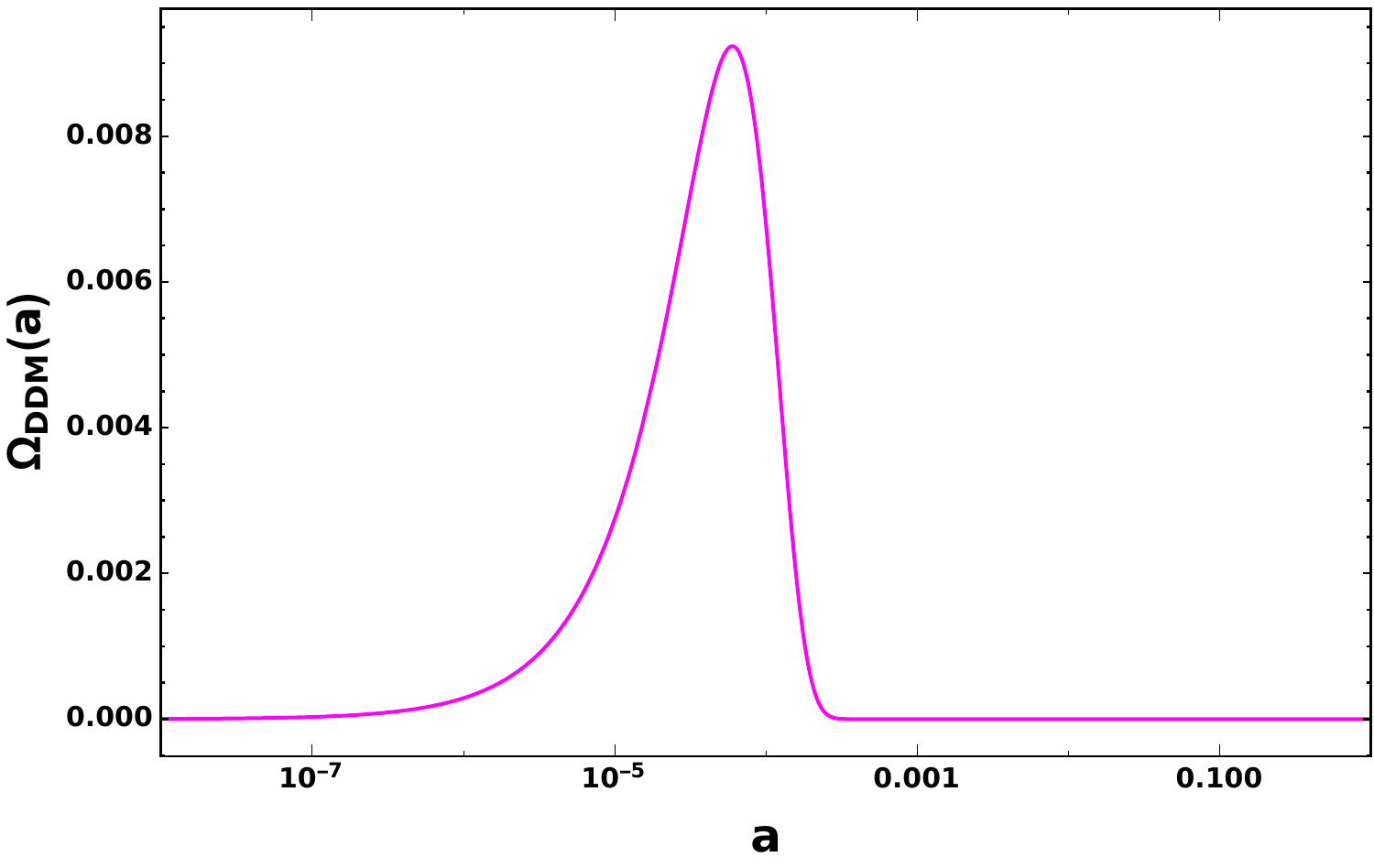}   
			\caption{ The left plot shows the contribution of different species in the energy budget in HTH model. The right plot is a zoom on only  $\Omega_{_{\rm DDM}}$. It is obvious that the $\Omega_{_{\rm DDM}}$ has a $\sim \%1$ contribution at most around the redshift $\sim 5\times 10^4$ and its contribution is very negligible before and after those redshifts.}
			\label{fig:omega}
		\end{center}
	\end{figure}

		Having obtained the numerical results, we can now verify the analytical expectations developed in Section \ref{pre-res}. For the best-fit parameter values in the HTH model, we obtain
		\begin{equation}
			r_s = 145.22\,\mathrm{Mpc},
		\end{equation}
		which is smaller than the $\Lambda$CDM prediction,
		\begin{equation}
			r_s = 147.09\,\mathrm{Mpc}.
		\end{equation}
		This reduction in the sound horizon, $\Delta r_s/r_s \simeq - 1.3\%$, shifts the CMB-inferred value of $H_0$ upward and illustrates how early-time modifications in the HTH scenario can help relieve the Hubble tension as we expected and discussed above. The contribution of $H(z)$ in this reduction is given by $\Delta r_s / r_s \simeq -\Omega_{_{\rm DDM}}/2$ while we assumed a constant $\Omega_{_{\rm DDM}}$ for the whole redshifts before the decaying to matter\footnote{This approximation is good since the very high redshifts have not any efficient contributions.}. The maximum of $\Omega_{_{\rm DDM}}\sim 0.01$ around the redshift $z=10^{5}$ which is shown in the Figure \ref{fig:omega}.  So $\Delta r_s / r_s \simeq -\Omega_{_{\rm DDM}}/2\sim \%1$ which is what we needed to address the Hubble tension. However there is a correction due to the sound speed $c_s$. This part is very negligible since the (additional) injected photons in HTH model is suppressed by $\Delta\rho_\gamma/\rho_\gamma\sim 10^{-6}$ and consequently, $\frac{\Delta c_s}{c_s} 
		\;\sim\; \frac{1}{2}\frac{\Delta\rho_\gamma/\rho_\gamma}{1+R} 
		\;\sim\; \mathcal{O}(10^{-7})$ and consequently $\Delta r_s/r_s\sim 10^{-7}$. So the main reason that HTH could resolve the Hubble tension is due to the modification of the Hubble parameter at early times.

		It is also worth make a comment on the smallness of injected photons' effects on the CMB distortions. We can estimate the fractional density of the injected photons at $a\simeq 10^{-5}$ (where the maximum of $\Omega_{_{\rm DDM}}$ occurs) as
		\begin{equation}
			\frac{\Delta\rho_\gamma}{\rho_\gamma}
			= \frac{\rho_{\rm DDM}(a_{\rm decay})}{\rho_\gamma(a_{\rm decay})} 
			\simeq \frac{\Omega_{\rm DDM}^{(0)}}{\Omega_\gamma^{(0)}}\,a_{\rm decay} \simeq 10^{-6},
			\label{eq:energy_injection}
		\end{equation}
		where for the last equality we have used the results reported in Table \ref{tab-best}. The $\mu$-distortion is given by\footnote{Note that this relation is assumed the single-epoch approximation which seems good enough for our purposes.}
		\begin{equation}
			\mu \;\approx\; 1.4\;\frac{\Delta\rho_\gamma}{\rho_\gamma}\;J(z_{\rm decay}),
			\qquad 
			J(z) \;\approx\; 1 - \exp\!\left[-\left(\frac{z}{z_{\rm dc}}\right)^{5/2}\right],
			\label{eq:mu_distortion}
		\end{equation}
		where $z_{\rm dc}$ is the double-Compton thermalization redshift \cite{Sunyaev:1970,Hu:1994bz}. For $z_{\rm dc}\sim 2\times 10^6$ and $z_{\rm decay}\sim 10^{5}$ the $\mu \sim 10^{-9}$. This $\mu$-value is far below the  FIRAS bound $\mu < 9\times 10^{-5}$ \cite{Chluba:2012we,Chluba:2011hw,Fixsen:1996nj}. The $y$-distortion is also very small since the decay is completed below $z \sim 5\times 10^4$. So it seems the HTH model can pass the $\mu$- and $y$-distortions' constraint. A precise quantification of these effects is left for future work.

	\section{Summary} \label{sec:summary}                                       
	In this work, we have explored the cosmological implications of the Hadrosymmetric Twin Higgs (HTH) model, 
	a realization of the twin Higgs framework in which the twin sector contains hadronic states but no additional light degrees of freedom. 
	In contrast to the Mirror Twin Higgs (MTH) model, where the discrete $\mathbb{Z}2$ symmetry enforces a full copy of the SM spectrum, 
	the HTH scenario features a hard breaking of this symmetry. As a result, 
	the twin sector is not a perfect replica of the SM but instead contains only a mirror copy of the SM hadrons. 
	In particular, it includes all three generations of twin quarks while excluding light states such as twin photons and twin neutrinos. 
	This structural difference makes the HTH model free from  $\Delta N_{\rm eff}$ constraints and highly elusive to collider searches, 
	but potentially relevant in cosmology.

	To study the model's impact on cosmological observables, we implemented the HTH sector in the Boltzmann solver \texttt{CLASS} and performed a Monte Carlo Markov Chain analysis using \texttt{MONTEPYTHON-v3}, confronting the model with Planck 2018 CMB data and local measurements of the Hubble constant. Following a phenomenological approach, we modeled the twin-sector states as an effective decaying dark matter component, allowing for interactions that produce visible photons.

	Our results show that the HTH framework can  alleviate the persistent tensions in modern cosmology. In particular, the Hubble tension is reduced from over $4\sigma$ in $\Lambda$CDM to approximately $2.5\sigma$, while the $\sigma_8$ discrepancy is simultaneously mitigated. Including the local measurement of $H_0$ further restricts the parameter space, leading to a scenario (HTH*) in which the tension in $\sigma_8$ is effectively removed.

	These findings highlight that the HTH model, while primarily designed to address naturalness in particle physics, also offers a viable mechanism to reconcile current cosmological observations. This underscores the potential of connecting high-energy theoretical models with precision cosmology and motivates further studies of twin-sector phenomenology in both particle physics and cosmology.
	\section*{Acknowledgments} \label{sec:ack}                                            
	ZD is supported the Korea Institute for Advanced Study grant no.6G097301 and 
	also acknowledges from Iran Science Elites Federation under grant number M401543.

	\bibliographystyle{JHEP}                                                                                      
	\bibliography{HTH}

\end{document}